\documentclass[aps,prd,nofootinbib,12pt]{revtex4}
\usepackage{graphics}
\usepackage{graphicx}
\usepackage{amsmath,amssymb,mathrsfs}
\usepackage{epsf}
\usepackage{slashed}
\usepackage{epsfig}
\usepackage{xcolor}
\allowdisplaybreaks

\newcounter{fig}   \newcommand{\lbfig}[1]{\refstepcounter{fig}
\label{#1} }

\newcommand{\bea}{\begin{eqnarray}}
\newcommand{\eea}{\end{eqnarray}}
\newcommand{\be}{\begin{equation}}
\newcommand{\ee}{\end{equation}}

\newcommand{\re}[1]{(\ref{#1})}

\newcommand{\eqn}{\begin{eqnarray}}
\newcommand{\eqnx}{\end{eqnarray}}

\begin{document}

\title{Q-balls in the $U(1)$ gauged Friedberg-Lee-Sirlin model}
\author{V.~Loiko}
\affiliation{ Department of Theoretical Physics and Astrophysics,
Belarusian State University, Minsk 220004, Belarus}
\author{Ya.~Shnir}
\affiliation{BLTP, JINR, Dubna 141980, Moscow Region, Russia
}

\begin{abstract}
We consider the $U(1)$ gauged two-component
Friedberg-Lee-Sirlin model in 3+1 dimensional Minkowski spacetime,
which supports non-topological soliton configurations.
Here we found families of  axially-symmetric spinning gauged
Q-balls, which possess both electric and magnetic fields.
The coupling to the gauge sector gives rise to a new branch of solutions,
which represent the soliton configuration coupled to a circular magnetic flux.
Further, in superconducting phase this branch is linked to  vorton type solutions
which represent a vortex encircling the soliton.
We discuss properties of these solutions and investigate their domains of existence.
\end{abstract}
\maketitle

\section{Introduction}
For a long time now, much attention has been paid to the soliton solutions of various
classical field theories. Solitons arise in various areas of theoretical and mathematical physics.
These spatially localized field configurations are widely used in many different contexts in
several directions including condensed matter physics, cosmology,
classical and quantum field theories, nuclear physics and other disciplines.
In many cases existence of the solitons is related with topological properties of the system,
their stability is secured by the conservation of the topological number, see \cite{Manton:2004tk}.
There are soliton configurations of another type, so called non-topological solitons that appear
as global minima in the corresponding classical action, see e.g \cite{Lee:1991ax,Shnir2018}.
A remarkable class of non-topological solitons commonly referred to as Q-balls,
exist in the field models possessing an unbroken, continuous global symmetry \cite{Rosen,Friedberg:1976me,Coleman:1985ki}.
These configurations carry a Noether charge associated with this symmetry,
they are time-dependent solitons with a stationary oscillating internal phase.

Configurations of this type were introduced by Rosen in 1968 \cite{Rosen}, later they were
revisited by Friedberg, Lee and Sirlin in two-component model with symmetry breaking potential
\cite{Friedberg:1976me}. In 1985 Coleman found another realization of Q-balls
considering a single complex scalar field
in a model with a non-renormalizable self-interaction potential \cite{Coleman:1985ki}.

The Friedberg-Lee-Sirlin model provides an interesting example of Q-ball solutions in a simple
renormalizable scalar theory with minimal interaction and symmetry breaking potential.
In this model the complex scalar becomes massive due to the coupling
with the real scalar field, since the latter has a finite vacuum expection
value generated via a symmetry breaking potential. Interestingly, the Q-ball solutions of that
model exist also in the limit of vanishing potential \cite{Levin:2010gp,Loiko:2018mhb}. In such a case
the real component of the coupled system becomes massless, it possess Coulomb-like asymptotic tail,
the configuration is stabilized by the gradient terms in the energy functional.

There has been a lot of interest in recent years in various aspects of  Q-balls.
In particular it was found that similar non-topological solitons appear in the
curved space-time, the gravitational interaction may lead to gravitational collapse of the
scalar field into the boson stars,
which represent compact, stationary spinning configurations with a harmonic time dependence
\cite{Kaup:1968zz,Ruffini:1969qy}. Certain types of  boson stars with appropriate non-linear self-interaction
are linked to the corresponding flat space solutions, which represent Q-balls
\cite{Friedberg:1986tp,Friedberg:1986tq,Kleihaus:2005me,Kleihaus:2007vk,Brihaye:2008cg,Kunz:2019sgn}.
It was suggested that these mini-boson stars  may contribute to various scenario of
the evolution of the early Universe \cite{Friedberg:1986tq,Jetzer:1991jr,Lee:1986ts}.
Further, it was argued that these Q-balls may play an essential role in baryogenesis via the Affleck-Dine
mechanism \cite{Affleck}, they also were considered as candidates for dark matter \cite{Kusenko:1997si}.

Notably, Q-ball configurations in the $U(1)$-gauged model of complex scalar field
with minimal electromagnetic coupling was considered already in the second of the pioneering papers of
Rosen \cite{Rosen}. Although Coleman expressed his doubts about possible existence of gauged Q balls
\cite{Coleman:1985ki}, existence of the corresponding solitons was confirmed by various authors
\cite{Lee:1988ag,Lee:1991bn,Kusenko:1997vi,Anagnostopoulos:2001dh,Gulamov:2015fya,Gulamov:2013cra,Panin:2016ooo,Loginov:2019sqf}.
Further, a possibility of generation of the magnetic field
by the angularly excited Q-balls was discussed in \cite{Shiromizu:1998eh}.

Indeed, in the simplest case the Q-balls are spherically symmetric, however there are
generalized spinning axially symmetric solutions with non-zero angular momentum
\cite{Kleihaus:2005me,Volkov:2002aj,Radu:2008pp}. The energy and the charge density distributions
of these rotating Q-balls represent a torus. An interesting aspect for such Q-balls is that there are two different types
of the axially-symmetric solutions with opposite parity \cite{Volkov:2002aj,Kleihaus:2005me,Kleihaus:2007vk,Brihaye:2008cg}.

Whereas various spherically symmetric $U(1)$-gauged Q-balls were investigated before, little is known about the
properties of the corresponding axially symmetric configurations, which possess both electric and magnetic field.
The main purpose of this work is to extend the consideration of
papers \cite{Lee:1988ag,Lee:1991bn,Kusenko:1997vi,Anagnostopoulos:2001dh,Gulamov:2015fya,Gulamov:2013cra,Panin:2016ooo}
by constructing new families of axially-symmetric stationary rotating
Q-balls in the $U(1)$-gauged Friedberg-Lee-Sirlin model  and investigate dynamical properties of the corresponding
configurations. We found that these solutions possess new properties, which are different from those
of the spherically symmetric
gauged Q-balls, featuring interesting pattern of generation of a toroidal magnetic field. Interestingly, new branch
of magnetic Q-balls arise, it corresponds to the non-topological soliton encircled by magnetic vortex. Further, we
observe that strong magnetic field may destroy the supeconducting phase in some region inside the Q-ball. The corresponding
configurations actually represent the vortons \cite{Davis:1988ij,Garaud:2013iba}, circular magnetic vortices
stabilized by the angular momentum of the stationary spinning soliton.

The paper is organized as follows: in the next section we discuss the $U(1)$-gauged
Friedberg-Lee-Sirlin model, the
axially-symmetric ansatz which we apply to parameterize
the action, the boundary conditions imposed to get regular solution and establish the field equations.
The numerical results are presented in
section 3, where we investigate  properties of these gauged
spinning Q-balls and determine their domains of existence.
We give our conclusions and remarks in the final section.

\section{The Model}
We consider the $U(1)$-gauged two-component
Friedberg-Lee-Sirlin model, which describes a coupled system of the
real self-interacting scalar field $\psi$ and a complex scalar field $\phi$, dynamically
coupled to an Abelian gauge field $A_\mu$. The corresponding Lagrangian density is
\be
L= -\frac14 F_{\mu\nu} F^{\mu\nu} + (\partial_\mu\psi)^2 + |D_\mu\phi|^2 - m^2 \psi^2|\phi|^2 - U(\psi) \, ,
\label{lag-fls}
\ee
where $D_\mu = \partial_\mu +igA_\mu$ denotes the covariant derivative,
the Abelian field strength tensor is $F_{\mu\nu}=\partial_\mu A_\nu-\partial_\nu A_\mu$
with electric components
$E_k=F_{k0}$ and magnetic components $B_k=\varepsilon_{kmn}F^{mn}$,
$g$ denotes the gauge coupling constant and $m$ is the coupling constant.
The potential of the real scalar field is
\be
U(\psi)= \mu (1-\psi^2)^2 \, ,
\label{Pot}
\ee
thus, $\psi \to 1$ in the vacuum, the $U(1)$ symmetry is broken inside the Q-ball and the gauge field $A_\mu$
becomes massive. In some sense, the gauged Q-ball behaves like a
superconductor \cite{Lee:1988ag}, here the component $\psi$ plays the role of the order parameter. The
normal phase corresponds to the case $\psi=0$, then the model \re{lag-fls} is reduced to the usual
scalar electrodynamics which does not support non-topological solitons.

Note that the model \re{lag-fls} may be considered as a truncated version of the  Witten's model
of superconducting cosmic strings with $U(1)\times U(1)$ local gauge invariance \cite{Witten:1984eb}.
Such a  theory supports stationary  vortex solutions \cite{Copeland:1987th,Davis:1988jp}, it also admits
the vortons, they represent the vortex rings stabilized by charge, current and
angular momentum \cite{Davis:1988ij,Garaud:2013iba}. As we will see, the gauged Friedberg-Lee-Sirlin model also
supports stationary superconducting circular loops with non-zero angular momentum.

The parameter $\mu$ defines the mass of the real component $\psi$,
the complex field $\phi$ becomes massive due to the coupling with its real partner.
The electromagnetic coupling also contributes decreasing the effective mass of the
field $\phi$,  as the coupling $g$ increases from zero.
In the limit of vanishing mass parameter $\mu \to 0$ but fixed
vacuum expectation value, the real scalar field
becomes massless and thus long-ranged.
Note that the complex component $\phi$ still acquires mass in this limit
due to the coupling with the Coulomb-like field $\psi$.

The model \re{lag-fls} is invariant under the usual local $U(1)$
transformations of the fields. The following conserved Noether current is associated with this symmetry,
\be
\label{Noether}
j_\mu = i(\phi D_\mu\phi^\ast-\phi^\ast D_\mu\phi) \, ,
\ee
with the corresponding charge $Q=\int{d^3x~ j^0}$. This current is a source in the corresponding
Euler-Lagrange equation for the electromagnetic field
\be
\partial^\mu F_{\mu\nu}= g j_\nu
\label{eq-em}
\ee

Variation of the Lagrangian \re{lag-fls} with respect to the
scalar fields leads to the equations of motion
\be
\label{scaleq}
\begin{split}
    \partial^\mu\partial_\mu \psi&=2\psi\left(m^2|\phi|^2+2\mu \left(1-\psi^2\right)\right),\\
    D^\mu D_\mu \phi&=m^2\psi^2\phi \, ,
\end{split}
\ee
Notably, the flat-space localized regular solutions
of the Friedberg-Lee-Sirlin model \re{lag-fls}
exist in the limit of vanishing scalar potential, $\mu \to 0$,
when the vacuum expectation value of the real component $\psi$ is kept non-zero
\cite{Levin:2010gp,Loiko:2018mhb}.
They represent gauged Q-balls with a long-range massless scalar component.

In the opposite limit, $\mu \to \infty$,
the real component of the model \re{lag-fls} trivializes, $\psi =1$,
and the massive complex field $\phi$ satisfies the usual equations of classical scalar electrodynamics.
Clearly, spatially localized stationary spinning solutions
of this equation do not exist in the flat space.

\subsection{Spinning axially-symmetric gauged Q-balls}
We are interested in stationary axially-symmetric solutions of the model \re{lag-fls}.
The usual ansatz for the scalar fields is
\be
\label{scalans}
\psi=X(r,\theta)\, , \qquad  \phi=Y(r,\theta)e^{i(\omega t+n\varphi)}\, ,
\ee
where $\omega$ is the spinning frequency of field, and
$n\in\mathbb{Z}$ is the azimuthal winding number. In the static gauge the electromagnetic field
can be parameterized as
\be
\label{Aans}
A_{\mu} dx^{\mu} =A_0(r,\theta)dt + A_\varphi(r,\theta) \sin\theta d\varphi
\ee
Substitution of this ansatz into the definition of the $U(1)$ charge $Q$ yields
\be
Q=\int d^3 x \left(gA_0+\omega \right) Y^2 \, ,
\ee
which is different from the particle number $N=\int d^3 x~Y^2 $.

Further, one can expect the usual angular frequency range, which for the ordinary Q-balls
in the decoupled limit $g=0$ is determined by the explicit structure of the potential, will be affected by
the electromagnetic interaction.

The stationary spinning axially symmetric configuration possess angular momentum
which is obtained from the  $T_\varphi^0$ component of the stress-energy tensor,
\be
J = \int d^3x~T_\varphi^0 = 4\pi \int\limits_0^\pi\int_0^\infty~r^2\sin\theta dr d\theta~\biggl\{
 (g A_0 + \omega ) (n + g A_\phi \sin\theta ) Y^2
+ J_{em}\biggr\} \, ,
\label{ang}
\ee
where the contributions of the electromagnetic field is
\be
J_{em}= \frac{1}{r^2} \partial_\theta A_0 \left(A_\phi\cos\theta + \sin\theta \partial_\theta A_\phi\right)
+\sin\theta  \partial_r A_\phi \partial_r A_0
\ee
The angular moment of the spinning gauged Q-ball is quantized in the units of the electric charge of the configuration,
$J=nQ$ \cite{Radu:2008pp}.

The total energy of the system becomes
\be
\begin{split}
E&=2\pi \int\limits_0^\pi\int_0^\infty~r^2\sin\theta dr d\theta~\biggl\{
X_r^2+Y_r^2 + \frac{X_\theta^2}{r^2} + \frac{Y_\theta^2}{r^2} +
\frac{1}{r^2}\left(g A_\phi +\frac{n}{\sin \theta}\right)^2 Y^2\\
&+ (gA_0 + \omega)^2 Y^2 + \mu(1-X^2)^2 + m^2 X^2 Y^2 + E_{em}\biggr\} \, ,
\end{split}
\label{energy}
\ee
where $X_{r,\theta}\equiv\partial_{r,\theta}X$, $Y_{r,\theta}\equiv\partial_{r,\theta}Y$,
and the electromagnetic energy density is
$$
E_{em}= \frac12\left\{ (\partial_r A_0)^2 + \frac{1}{r^2}(\partial_\theta A_0)^2
+\frac{1}{r^2}(\partial_r A_\phi)^2 + \frac{1}{r^4\sin^2\theta}
\left[\partial_\theta(A_\phi \sin\theta)\right]^2\right\}
$$

The corresponding field equations are
\be
\begin{split}
&\left(\frac{\partial^2}{\partial r^2} +
\frac{2}{r}\frac{\partial}{\partial r} + \frac{1}{r^2} \frac{\partial^2}{\partial \theta^2}
+ \frac{\cos \theta}{r^2 \sin\theta}\frac{\partial}{\partial \theta}
+ 2 \mu^2 (1-X^2) - m^2 Y^2
 \right) X = 0\, ;\\
&
\left(\frac{\partial^2}{\partial r^2} +
\frac{2}{r}\frac{\partial}{\partial r} + \frac{1}{r^2} \frac{\partial^2}{\partial \theta^2}
+ \frac{\cos \theta}{r^2 \sin\theta}\frac{\partial}{\partial \theta}
-\frac{1}{r^2} \left(gA_\phi +\frac{n}{\sin\theta} \right)^2
+ (gA_0+\omega )^2 - m^2 X^2
 \right) Y = 0\, ;\\
&
\left(\frac{\partial^2}{\partial r^2} +
\frac{2}{r}\frac{\partial}{\partial r} + \frac{1}{r^2} \frac{\partial^2}{\partial \theta^2}
+ \frac{\cos \theta}{r^2 \sin\theta}\frac{\partial}{\partial \theta}
-\frac{1}{r^2\sin^2\theta} - 2 g^2 Y^2 \right) A_\phi = \frac{2n g}{\sin\theta} Y^2\, ;\\
&
\left(\frac{\partial^2}{\partial r^2} +
\frac{2}{r}\frac{\partial}{\partial r} + \frac{1}{r^2} \frac{\partial^2}{\partial \theta^2}
+ \frac{\cos \theta}{r^2 \sin\theta}\frac{\partial}{\partial \theta}
-2 g^2 Y^2\right) A_0 = 2g\omega Y^2 \, .
\end{split}
\label{eqs}
\ee
Here the last equation represents the Gauss law, it has to be considered as a constraint imposed on
the system. Setting $n=0$ reduces the equations \re{eqs} to the spherically symmetric gauged Q-balls, considered
in \cite{Lee:1991bn}. Here we mainly focus on the investigation of the axially symmetric solutions with $n=1$,
evidently the main difference is that these configurations possess both electric and magnetic fields. As
we will see, the presence of the magnetic field strongly affects the properties of the Q-balls.

Note that the structure of the system of equation \re{eqs} suggests that,
similar to the case of the spinning axially-symmetric Q-balls
\cite{Volkov:2002aj,Kleihaus:2005me,Kleihaus:2007vk,Radu:2008pp,Loiko:2018mhb}, the solutions of the
gauged Friedberg-Lee-Sirlin model \re{lag-fls} may be either symmetric with respect to reflections in the equatorial
plane, $\theta \to \pi -\theta$,  or antisymmetric.
Here we restrict our consideration to the case of symmetric parity-even solutions. Further, in our numerical calculations
we set $m=1$ retaining other parameters of the model \re{lag-fls}.

\subsection{Numerical scheme and the boundary conditions}

To find numerical solutions of the coupled partial differential equations \re{eqs}
we used the software package CADSOL based on the Newton-Raphson algorithm \cite{schoen}.
The numerical calculations are mainly performed on an equidistant grid
in spherical coordinates $r$ and $\theta$. Typical grids we used have sizes $70 \times 60$.
In our numerical scheme we map the infinite interval of the variable $r$ onto the compact radial coordinate
$x=\frac{r/r_0}{1+r/r_0}  \in [0:1]$. Here $r_0$ is a real scaling constant,
that is used to improve the accuracy of the numerical solution.
Typically it is taken as $r_0 = 2 - 6$. Estimated numerical errors are of order of $10^{-5}$.

The system \re{eqs} represents a set of four coupled elliptic partial differential
equations with mixed derivatives, to be solved numerically
subject to the appropriate boundary conditions. As usual,
they follow from the condition of regularity of the fields on
the symmetry axis and symmetry requirements, as well as
the condition of finiteness of the energy of the system.
Explicitly, at the origin we impose
\be
\partial_r X\bigl.\bigr|_{r=0,\theta}=0  \, , \qquad \partial_r Y\bigl.\bigr|_{r=0,\theta}=0
\, , \qquad \partial_r A_0\bigl.\bigr|_{r=0,\theta}=0
\, , \qquad \partial_r A_\phi\bigl.\bigr|_{r=0,\theta}=0
\ee
while the boundary conditions on spatial infinity are
\be
X\bigl.\bigr|_{r=\infty,\theta}=1  \, , \qquad Y\bigl.\bigr|_{r=\infty,\theta}=0
\, , \qquad  A_0\bigl.\bigr|_{r=\infty,\theta}=0
\, , \qquad A_\phi\bigl.\bigr|_{r=\infty,\theta}=0
\label{boundary_r}
\ee
Finally, to secure the condition of regularity on the symmetry axis we
impose there the boundary conditions
\be
\partial_\theta X\bigl.\bigr|_{\theta=0,\pi}=0  \, , \qquad Y\bigl.\bigr|_{\theta=0,\pi}=0 \, ,
\qquad  \partial_\theta A_0\bigl.\bigr|_{\theta=0,\pi}=0
\, , \qquad A_\phi\bigl.\bigr|_{\theta=0,\pi}=0
\ee

\section{Numerical results}

Spherically symmetric solutions of the gauged Friedberg-Lee-Sirlin model
have been studied before \cite{Lee:1991bn}. The general
pattern is that the $n=0$  gauged Q-balls exist in the restricted domain of values of the
parameters of the system, there is a critical solution with maximal
charge and energy. The repulsive electromagnetic interaction reduces the allowed range
of values of the angular frequency of the spinning gauged Q-ball, in the decoupled limit the ordinary
Friedberg-Lee-Sirlin Q-balls exist for all non-zero values of  scaled frequency
$\omega \in [0,\omega_\mathrm{max}=1]$.
Here the upper bound corresponds to the mass of the complex scalar field, as $\omega$ approaches the upper
bound, the size of the Q-ball is decreasing and the configuration tends to the perturbative spectrum of
linearized excitations.

\begin{figure}[hbt]
\begin{center}
\includegraphics[height=.34\textheight, angle=-90]{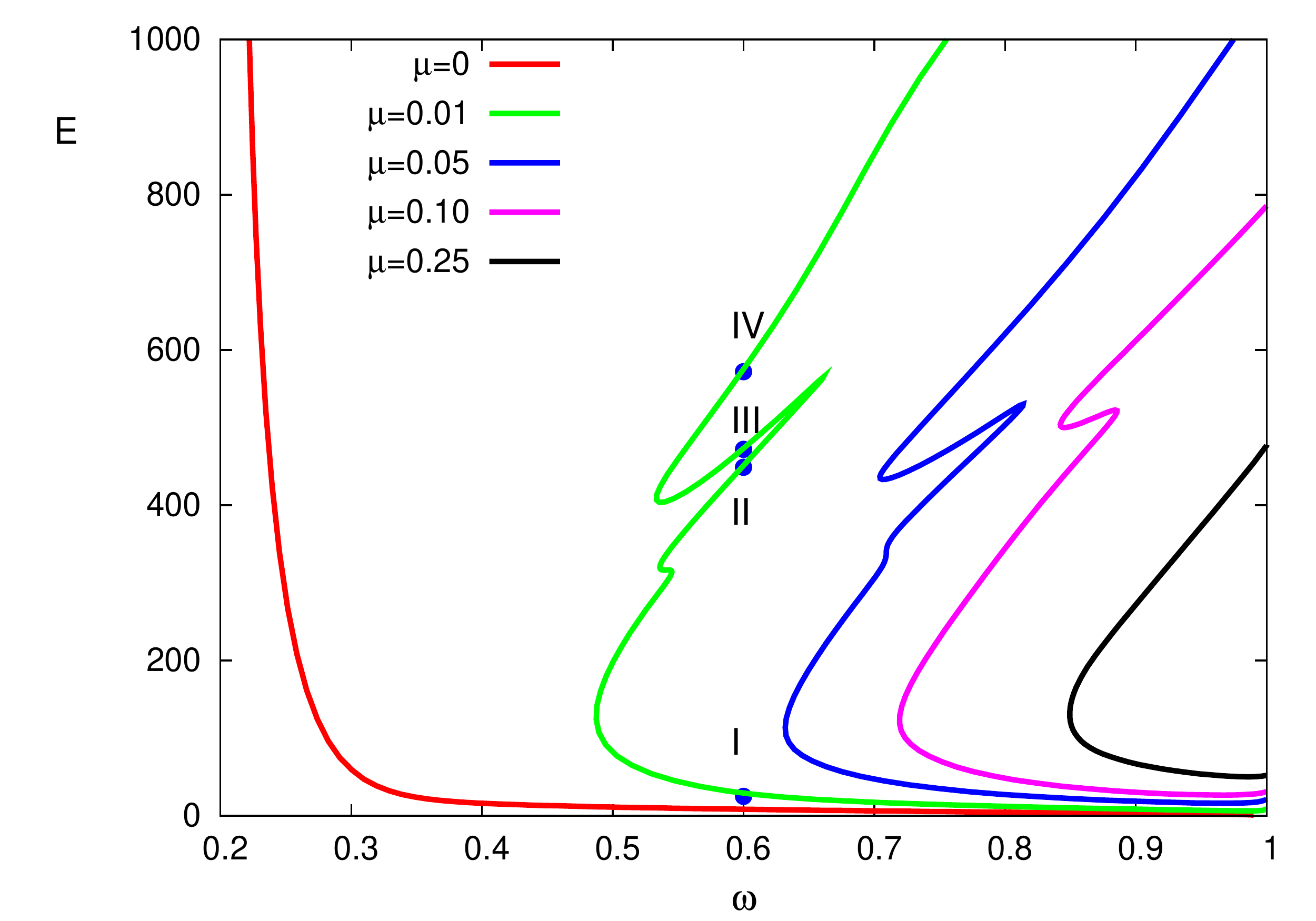}
\includegraphics[height=.34\textheight, angle=-90]{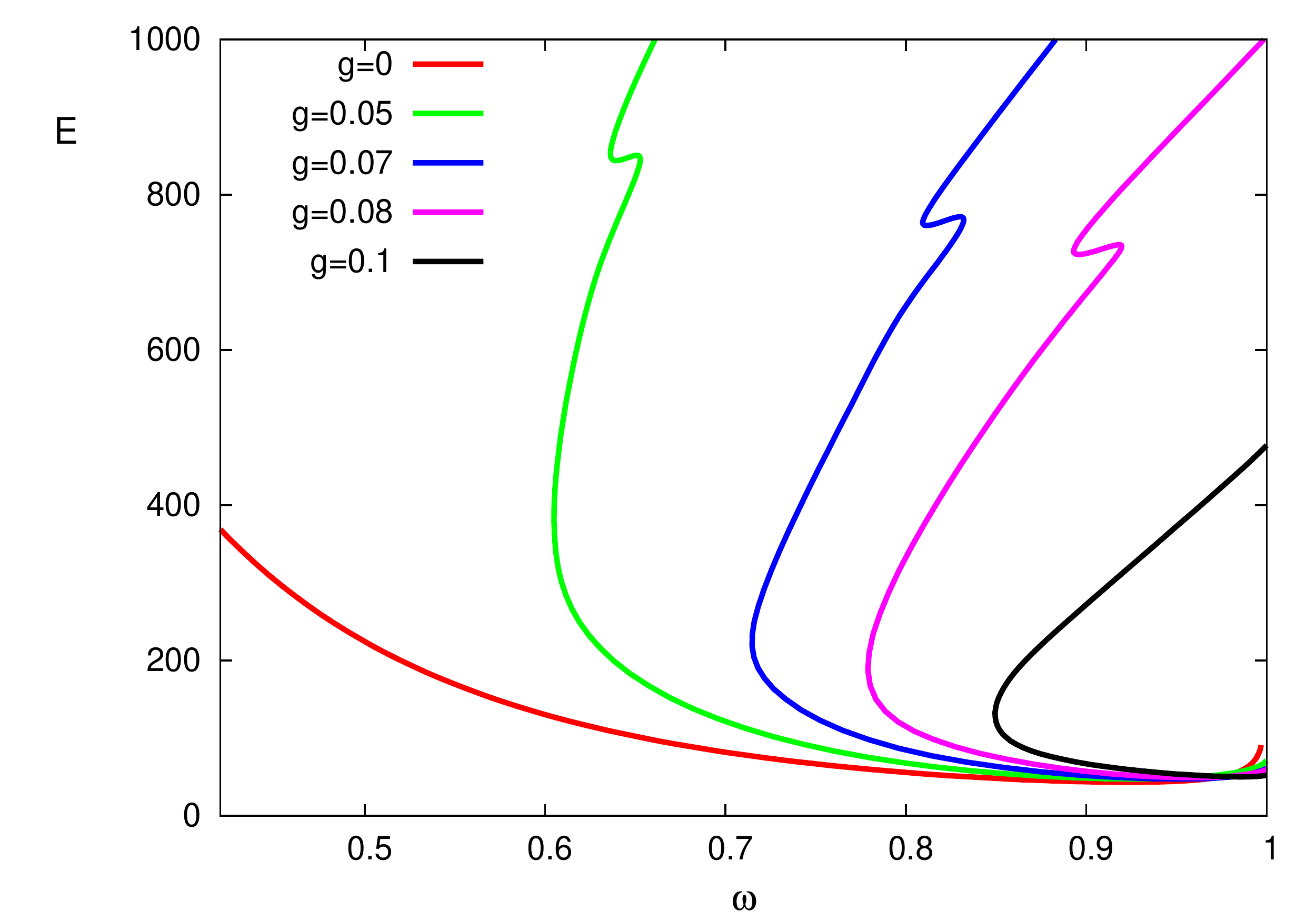}
\end{center}
\caption{\small
The total energy of the parity-even $n=1$ gauged Q-balls is shown as function of the
angular frequency $\omega$
for some set of values of mass parameter $\mu$ at $g=0.1$ (left plot) and
for some set of values of the gauge coupling $g$ at $\mu=0.25$ (right plot)
The numbers on the curves on the left plot correspond to the plots
in Figs.~\ref{E-3d},\ref{Em-3d} below.
}
\lbfig{E-omega}
\end{figure}

For axially symmetric  solutions of the gauged model \re{lag-fls} the frequency $\omega$
is also bounded from above.
As we shall see,  the electromagnetic coupling affects the lower critical value of the frequency, which is no
longer equal to zero.

To demonstrate the effects of electromagnetic interaction on the spinning solutions,
we exhibit in Fig.~\ref{E-omega} the total energy of the parity-even $n=1$ gauged Q-balls as
a function of the angular frequency $\omega$ at a given value of the gauge coupling $g=0.1$ and some set
of values of the mass parameter $\mu$ (left plot) and for some set of values of the gauge coupling at fixed
mass $\mu=0.25$ (right plot). First, we observe that the solutions exist as the gauge coupling remains relatively weak,
for $\mu=0.25$ the allowed range of values of the coupling is restricted as $g\le 0.15$.
Indeed, as the gauge coupling increases, the electrostatic repulsion becomes stronger than the scalar attraction
and localized solutions cease to exist. On the other hand, the critical value of the gauge coupling depends on the
value of the mass parameter, it increases as $\mu$ decreases.
The upper critical value of the angular frequency still remains
$\omega_\mathrm{max}=1$. Indeed, as $r\to \infty$ the system  \re{eqs} with the boundary
conditions \re{boundary_r} is approaching the usual Laplace equation for all components, the fields become oscillating
as $\omega > \omega_\mathrm{max}$. Evidently, such a system cannot
support localized solutions. However, both the energy and the charge of the gauged axially symmetric Q-balls do not diverge
as $\omega \to \omega_\mathrm{max} $, see Fig.~\ref{E-omega}.

The spinning gauged Q-balls smoothly arise as the angular frequency is decreasing below  $\omega_\mathrm{max}$, see
Fig.~\ref{E-omega}. Forming a branch of solution, which are similar to the ordinary Q-balls,
these configurations evolve smoothly as the angular frequency is decreasing. The solutions posses both electric and
magnetic field, which is generated by the Noether current $j_\mu$ \re{Noether}.
The corresponding toroidal magnetic field encircles the Q-ball, as seen in Fig.~\ref{mag-field}.
Further, the electric charge of the configuration is vanishing at the center of the spinning gauged Q-ball,
it is pushed outwards.

\begin{figure}[hbt]
\begin{center}
\includegraphics[height=.34\textheight, angle=0]{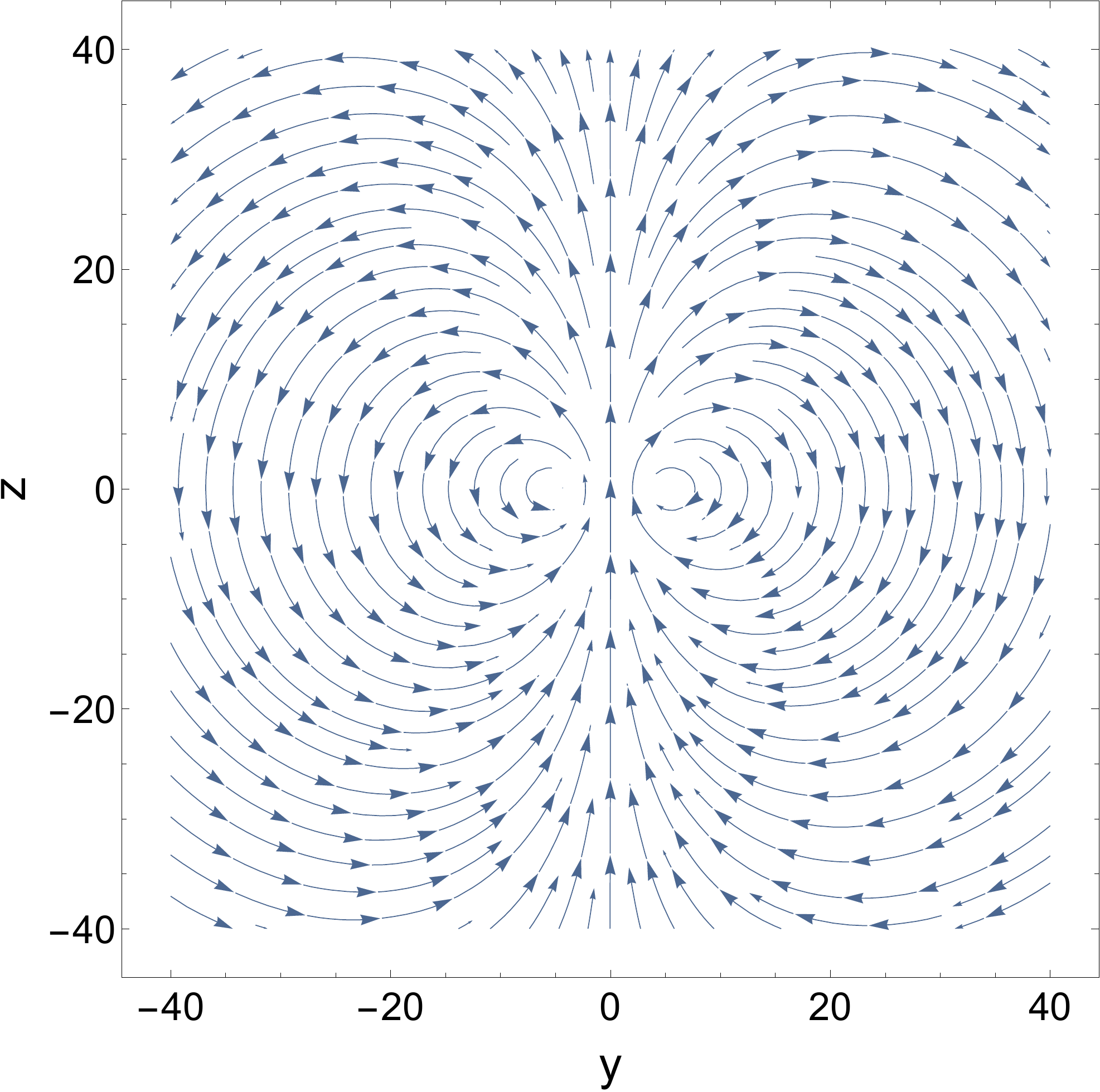}
\includegraphics[height=.34\textheight, angle=0]{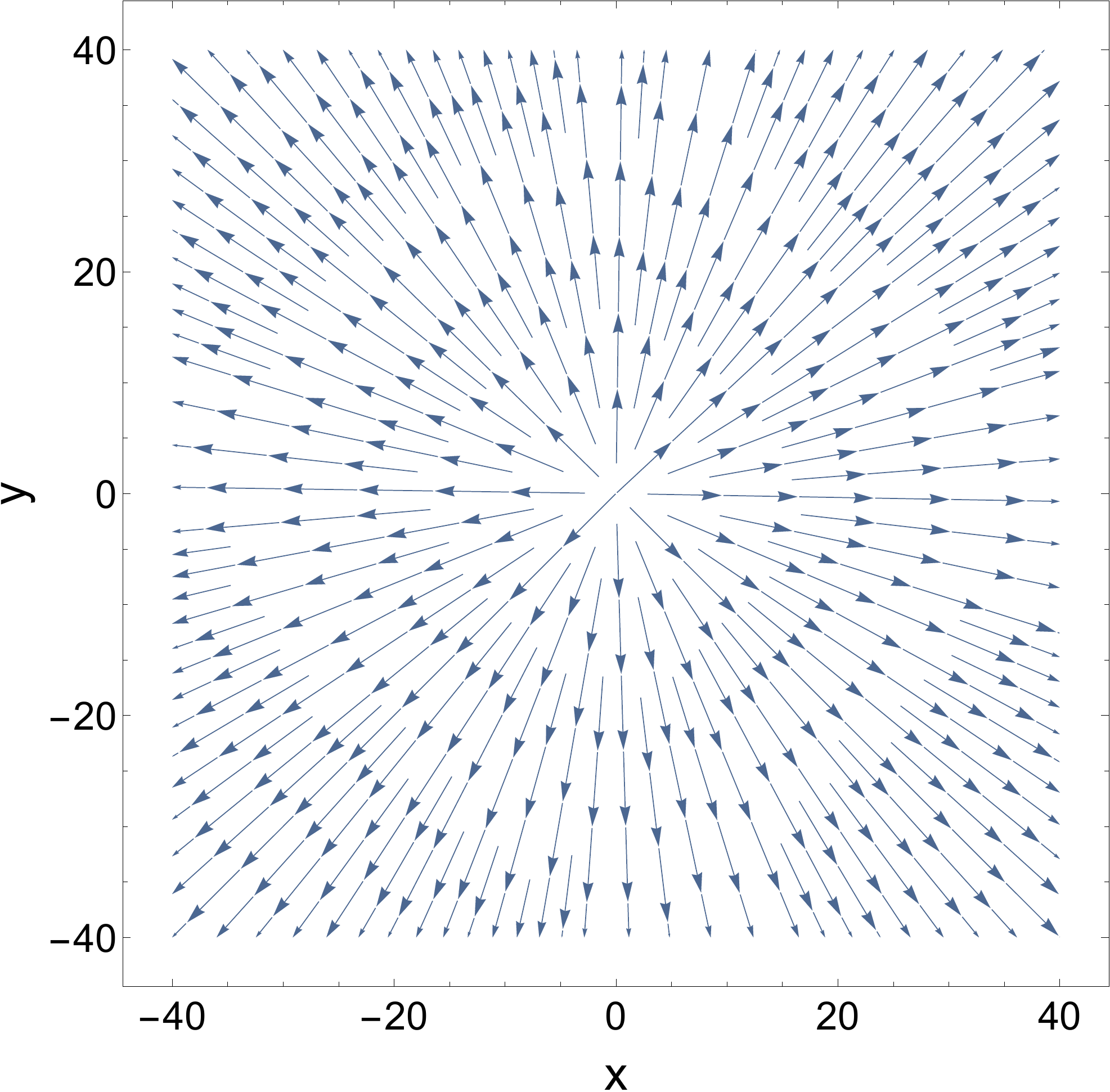}
\end{center}
\caption{\small
Magnetic field orientation of the gauged $n=1$ Q-ball at $g=0.1, \mu=0.25$ and $\omega=0.85$ (electric
branch); the magnetic flux in the $y-z$ plane (left plot) and in the $x-y$ plane (right plot).
}
\lbfig{mag-field}
\end{figure}

As the gauge coupling is small enough, the electromagnetic energy of the spinning Q-ball
remains smaller than the other contributions to the total energy \re{energy}, further, on that branch the contribution
of the electrostatic energy is much higher that the energy of the magnetic field of the configuration. We will refer
to that branch as "electric" one. Note that the $U(1)$ symmetry remains broken inside the Q-ball, so the configurations
remains in the "superconductive" phase.

The characteristic size of the gauged Q-ball increases as the angular frequency is decreasing along the electric branch,
hence both the current $j_\mu$ and associated magnetic field, become stronger. For some critical value of the frequency
$\omega_\mathrm{min}$ the value of the real component of the configuration on the $x-y$ plane
approaches zero, then the electromagnetic field becomes massless on a circle in the equatorial plane. As a result, the
energy of the magnetic field  becomes higher than the electrostatic energy of the spinning Q-ball.

\begin{figure}[hbt]
\begin{center}
\includegraphics[height=.34\textheight, angle=-90]{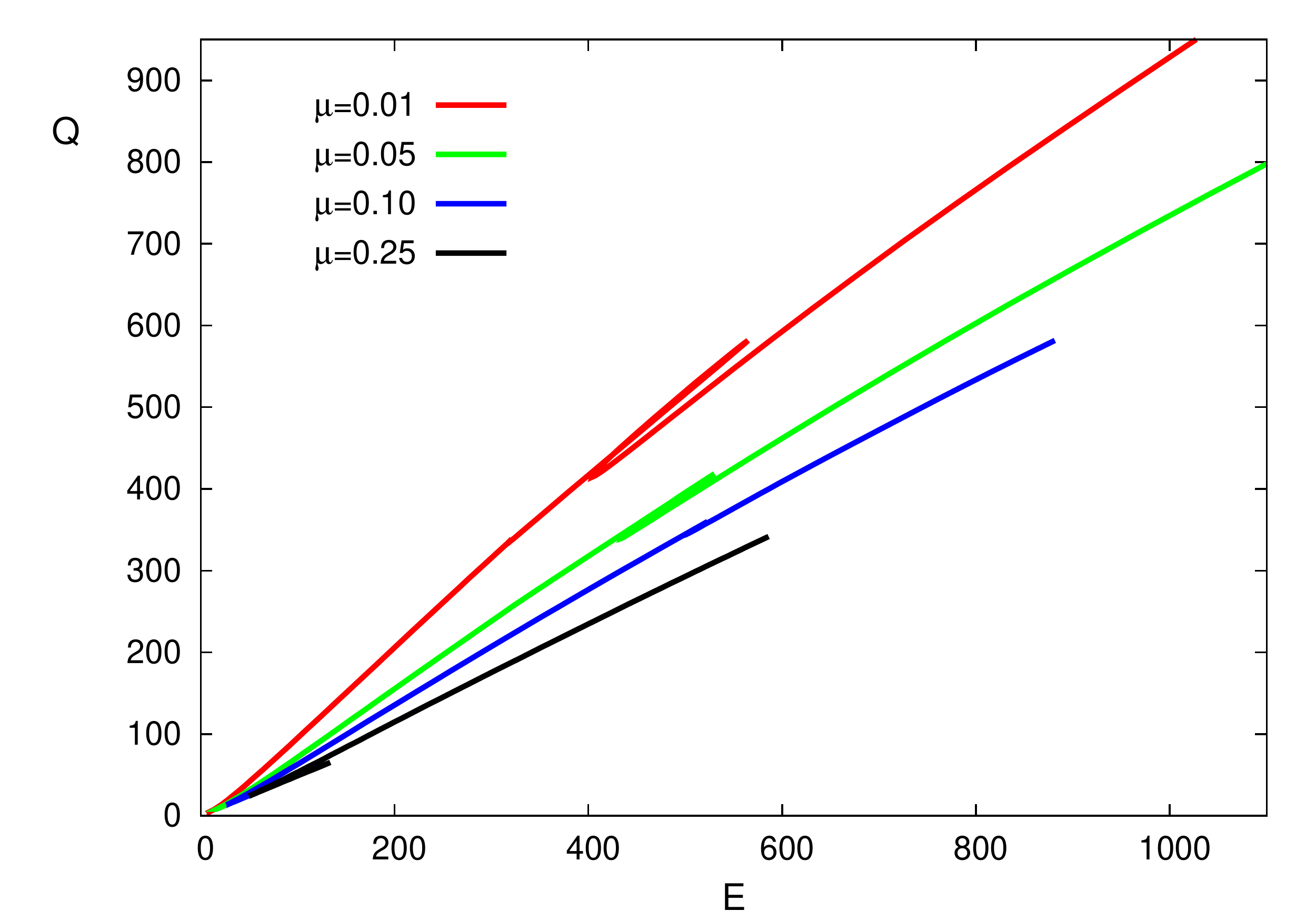}
\includegraphics[height=.34\textheight, angle=-90]{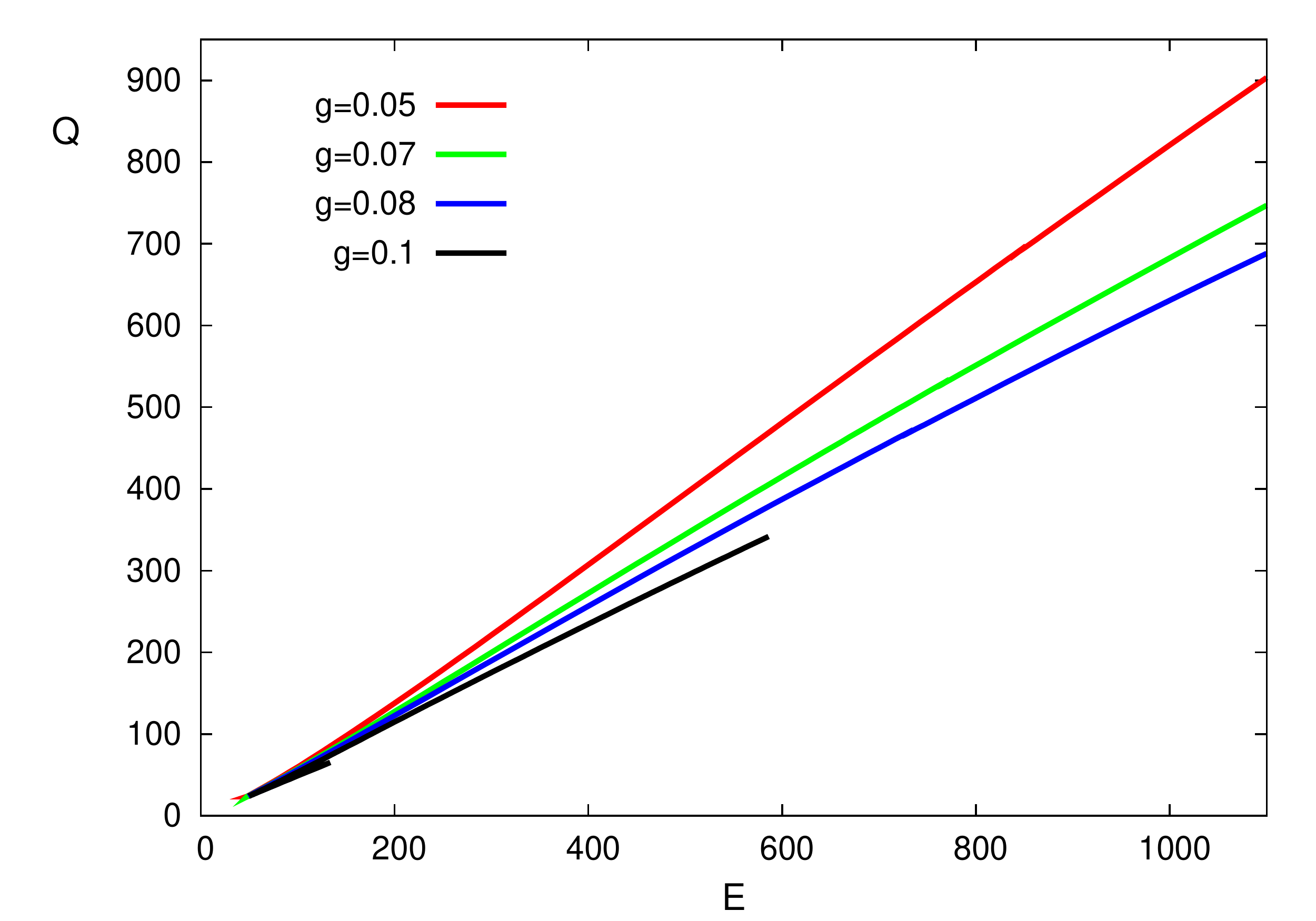}
\end{center}
\caption{\small
The total energy of the parity-even $n=1$ gauged Q-balls vs the
charge $Q$ for some set of values of mass parameter $\mu$ at $g=0.1$ (left plot) and
for some set of values of the gauge coupling $g$ at $\mu=0.25$ (right plot).
}
\lbfig{E-Q}
\end{figure}

As the angular frequency increases, the magnitude of the magnetic field increases significantly,
the second, magnetic branch of solutions extends forward and the energy of the
configuration increases rapidly, see Fig.~\ref{E-omega}. In other words, strong magnetic
field forms a domain of "normal" phase  inside of a superconductor.
Further evolution along this branch, in general is not monotonous, for small values of the mass
parameter $\mu$ additional branches may arise. However, in the presence of the gauge field
the angular frequency is not quite appropriate physical quantity, which is, however, a useful parameter
in our numerical simulations. Since for the gauged Q-balls the total energy and the charge decrease and increase
simultaneously, these quantity possess extrema at the same critical values of angular velocity, the relation holds
\cite{Gulamov:2013cra,Nugaev:2019vru}
$$
\frac{\partial E}{\partial Q}=\omega
$$
Thus, it is instructive to consider the curves of dependency  $E(Q)$, exhibited in
Fig.~\ref{E-Q}. The cusps on these curves occurs at the minimal values
of the charge $Q$, where the lower (electric) and upper (magnetic)
branch merge \cite{Tamaki:2010zz,Kleihaus:2011sx}.
Evidently, existence of two different solutions
with the same value of charge Q indicates that the more
energetic configurations on the upper branch are unstable, the cusps usually indicate the
boundary between the regions of stability of gauged Q-balls.
One can expect the magnetic branch could be unstable.

Our simulations show that along the magnetic branch a plateau of zero values of the real
component is formed, the domain of normal phase  further extends  as the frequency grows.
In Fig.~\ref{Profiles} we plotted the profiles of the field components $X(r,\pi/2)$, $Y(r,\pi/2)$ on the equatorial plane
for 4 different branches at the same values of the angular frequency
$\omega=0.60$ and $\mu=0.01$, the corresponding
configurations are labeled by the dots on the curves in Fig.\ref{E-omega}. Further, in Figs.~\ref{E-3d},\ref{Em-3d}
we exhibit the distributions of the energy density and the electromagnetic energy density of the corresponding
configurations as functions of the cylindrical coordinates  $\rho=r \sin\theta$ and $z=r\cos \theta$.

The magnetic branch exist for all non-zero values of the mass parameter $\mu$, it extends almost linearly
with $\omega$. Further increase of the frequency leads to expansion of the domain of normal phase
in which the real component of the configuration is trivial, $\psi=0$,
and both electromagnetic and complex field are massless. Thus, the magnetic field of the vortex
becomes stronger and the circular wall which separates the phases $\psi=0$ and $\xi=1$, approaches to the step function
as the angular frequency continues to increase.

\begin{figure}[h!]
\begin{center}
\includegraphics[height=.34\textheight,  angle =-90]{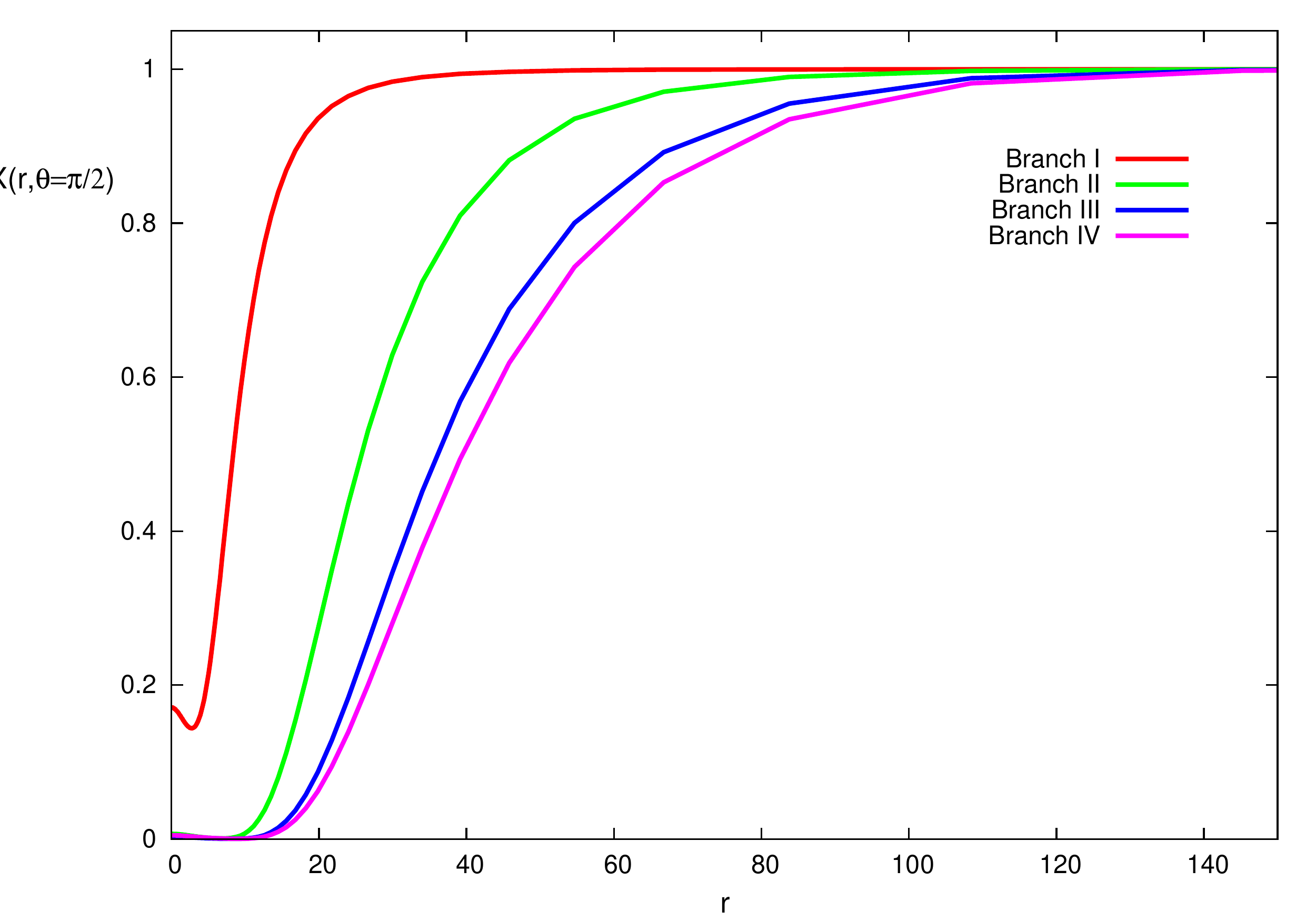}
\includegraphics[height=.34\textheight,  angle =-90]{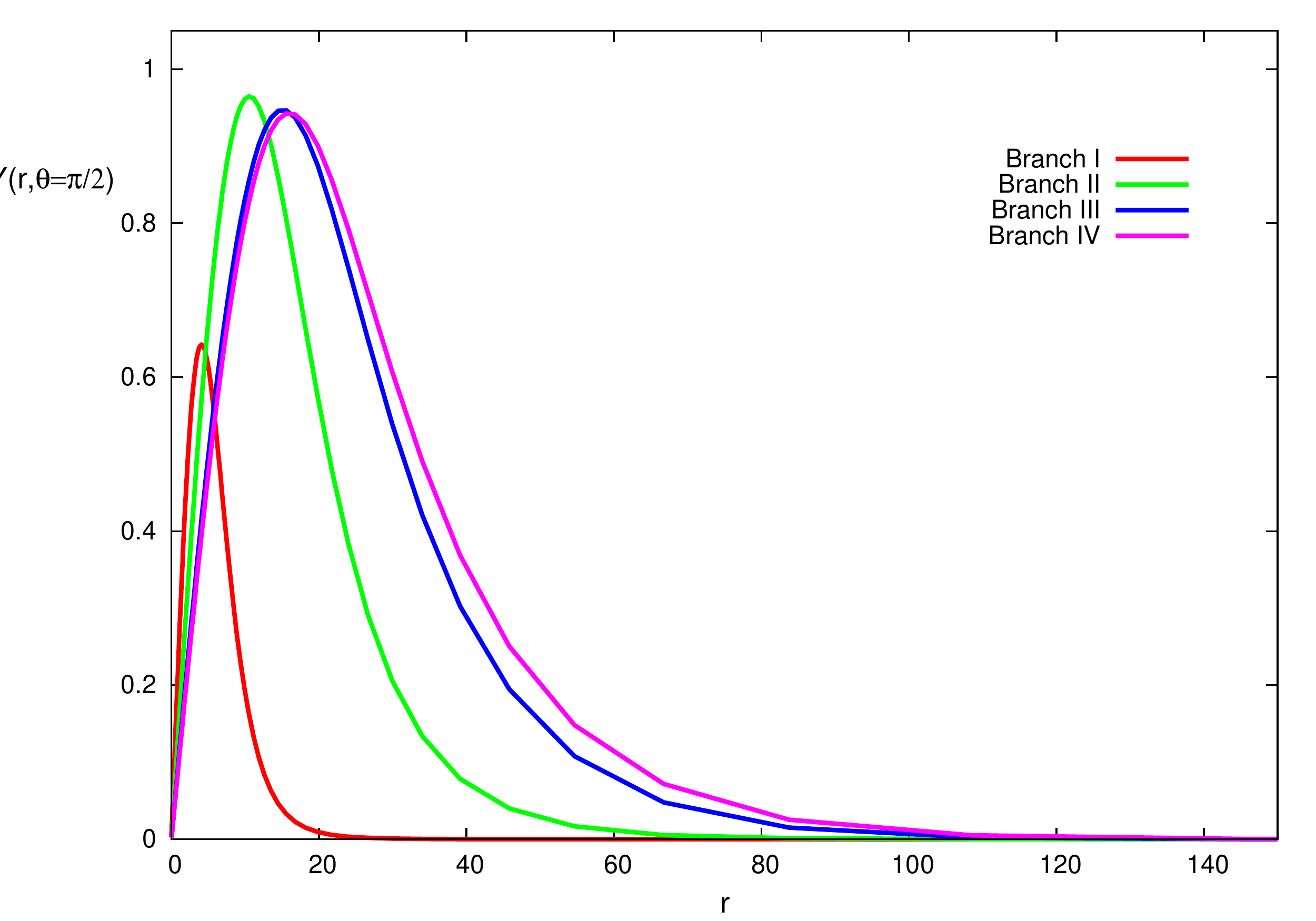}
\end{center}
\caption{\small
The profiles of the field components of the gauged Friedberg--Lee--Sirlin Q-balls  $X$ (left plot)
and $Y$ (right plot) at $\theta=\pi/2$
are plotted on four different branches, labeled by dots on the curves in
Fig.\ref{E-omega}, at $\omega=0.60$, $\mu=0.01$
and $g=0.1$.}
    \lbfig{Profiles}
\end{figure}

\begin{figure}[hbt]
    \begin{center}
    \includegraphics[width=.34\textwidth, angle=-90]{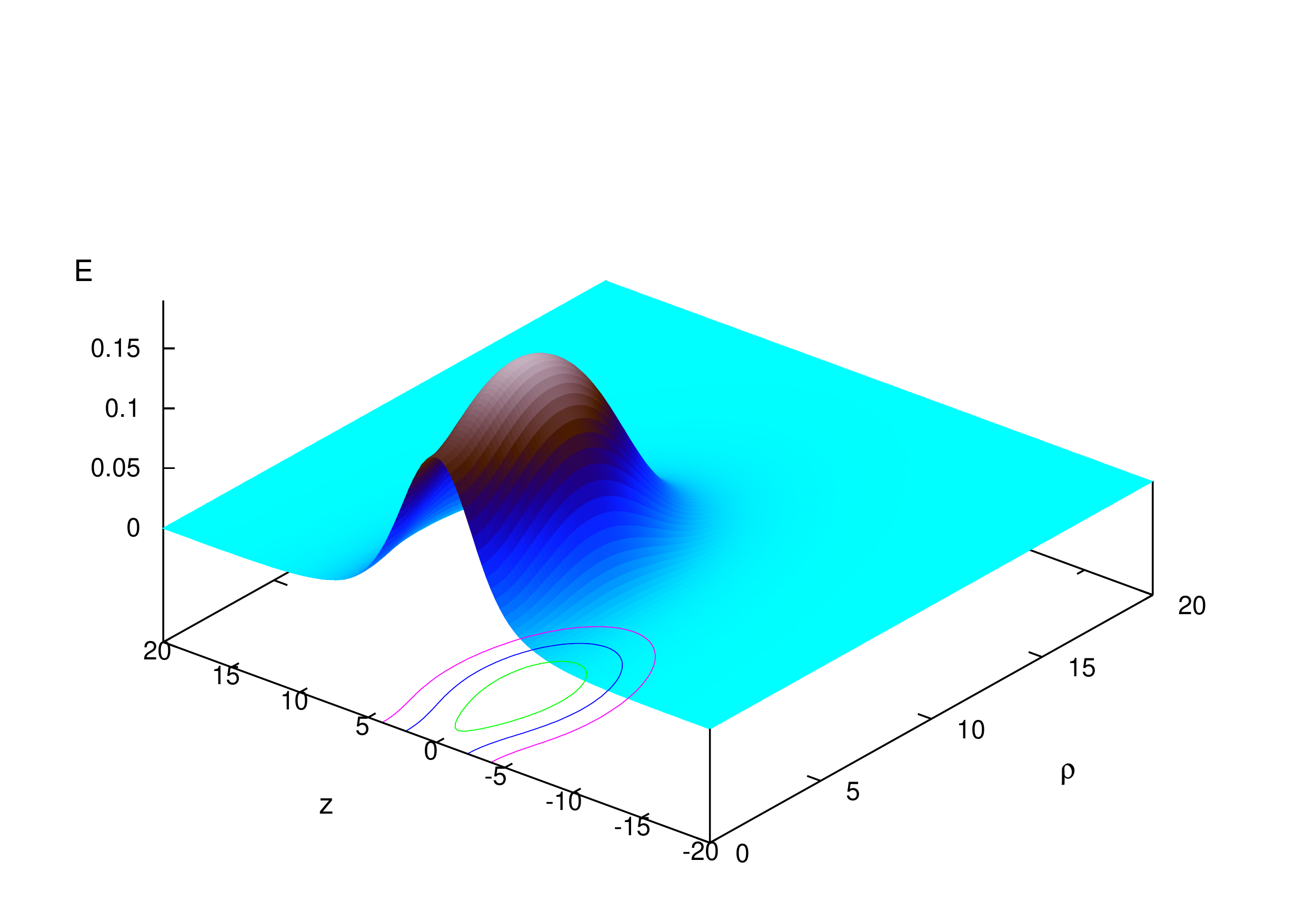}
    \includegraphics[width=.34\textwidth, angle=-90]{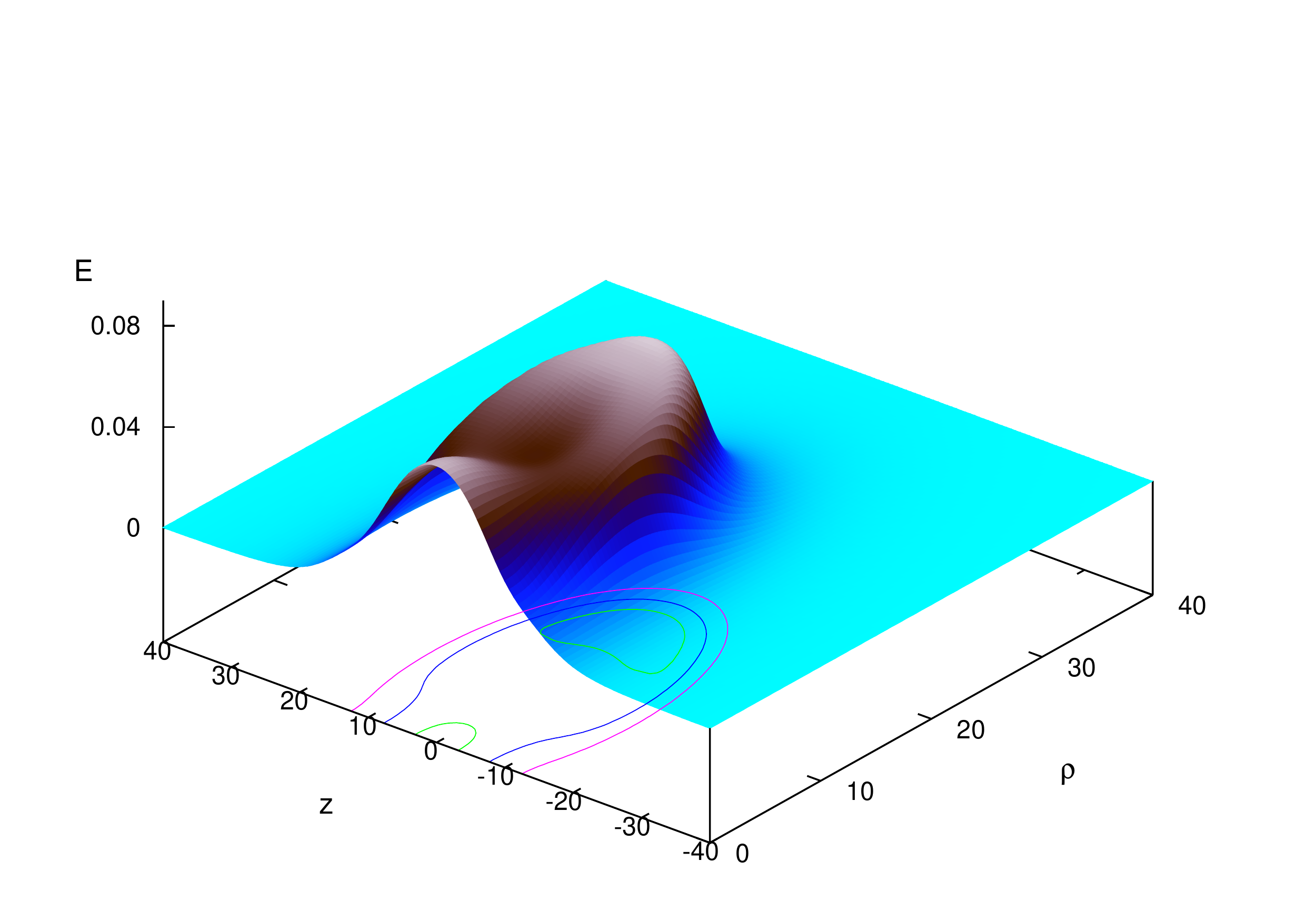}\\[18pt]
\hspace{0mm} I \hspace{67mm} II\hspace{27mm}
    \includegraphics[width=.34\textwidth, angle=-90]{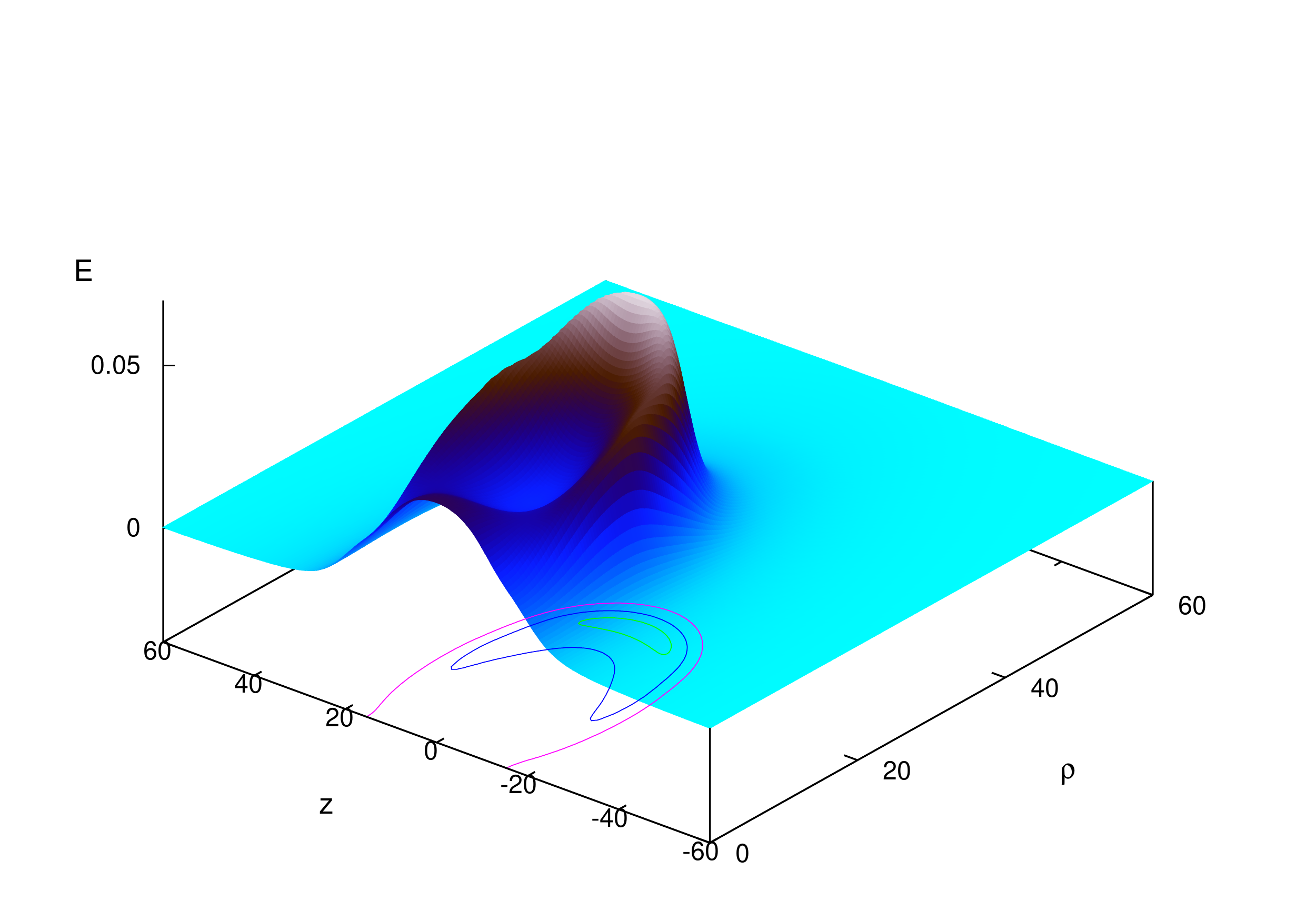}
    \includegraphics[width=.34\textwidth, angle=-90]{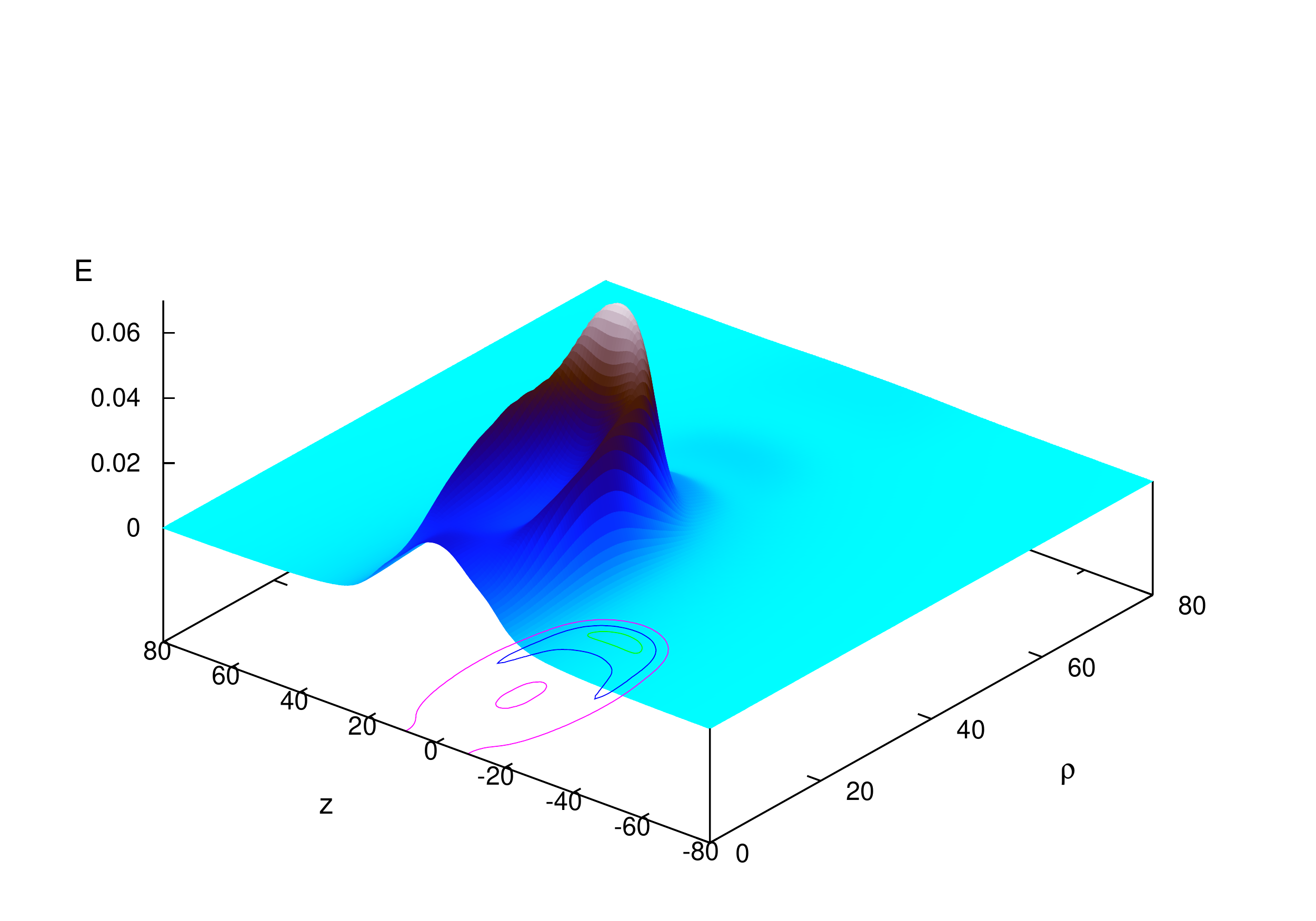}\\[18pt]
    \hspace{0mm}  III\hspace{67mm} IV\hspace{27mm}\\
    \end{center}
    \caption{\small
        The distributions of the total energy density of the gauged parity-even Q-balls on
        four different branches at $n=1, \mu = 0.01$ and $\omega=0.60$ are shown as functions of the coordinates $\rho=r \sin\theta$
        and $z=r\cos \theta$. }
    \lbfig{E-3d}
\end{figure}

\begin{figure}[hbt]
    \begin{center}
    \includegraphics[width=.34\textwidth, angle=-90]{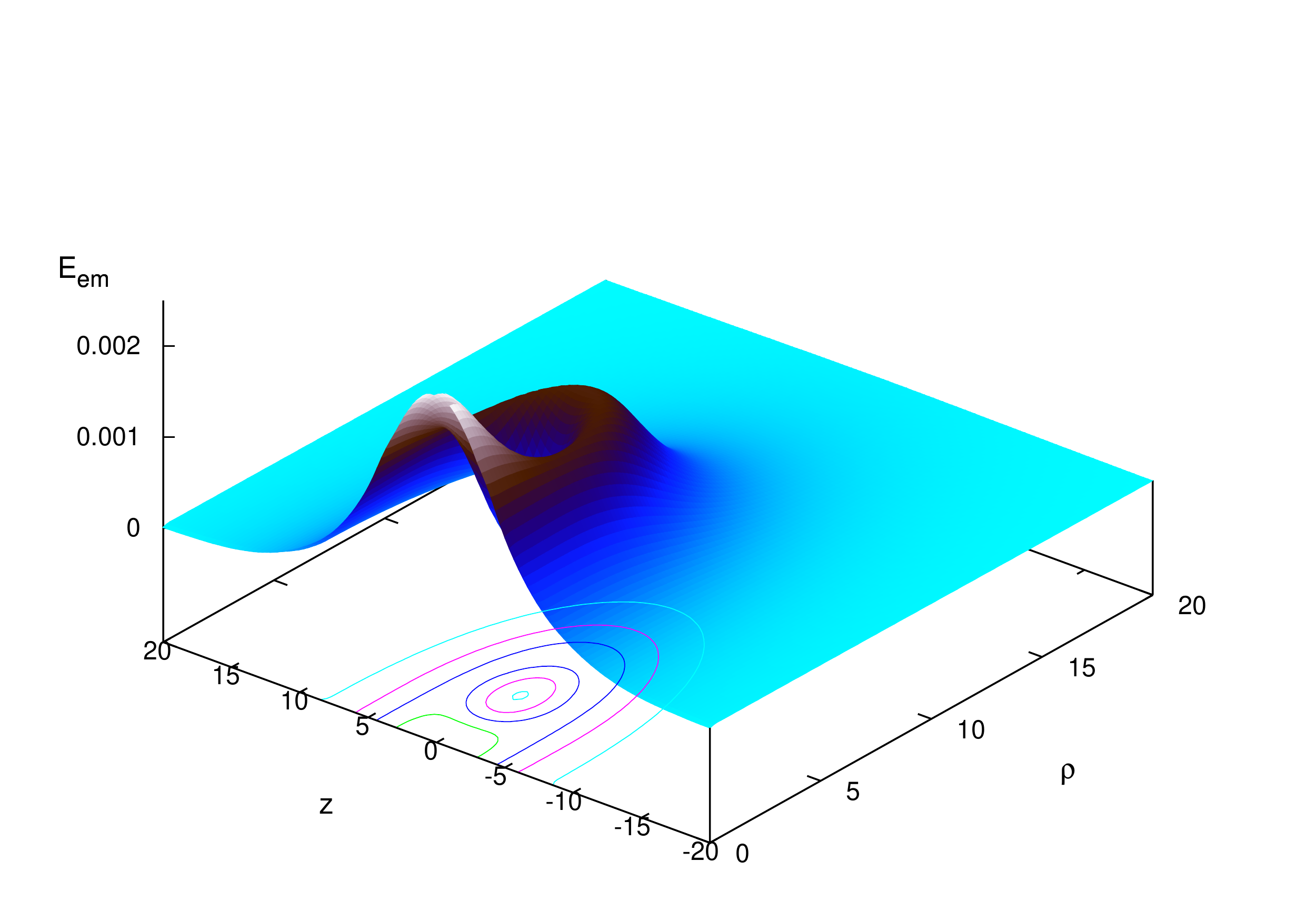}
    \includegraphics[width=.34\textwidth, angle=-90]{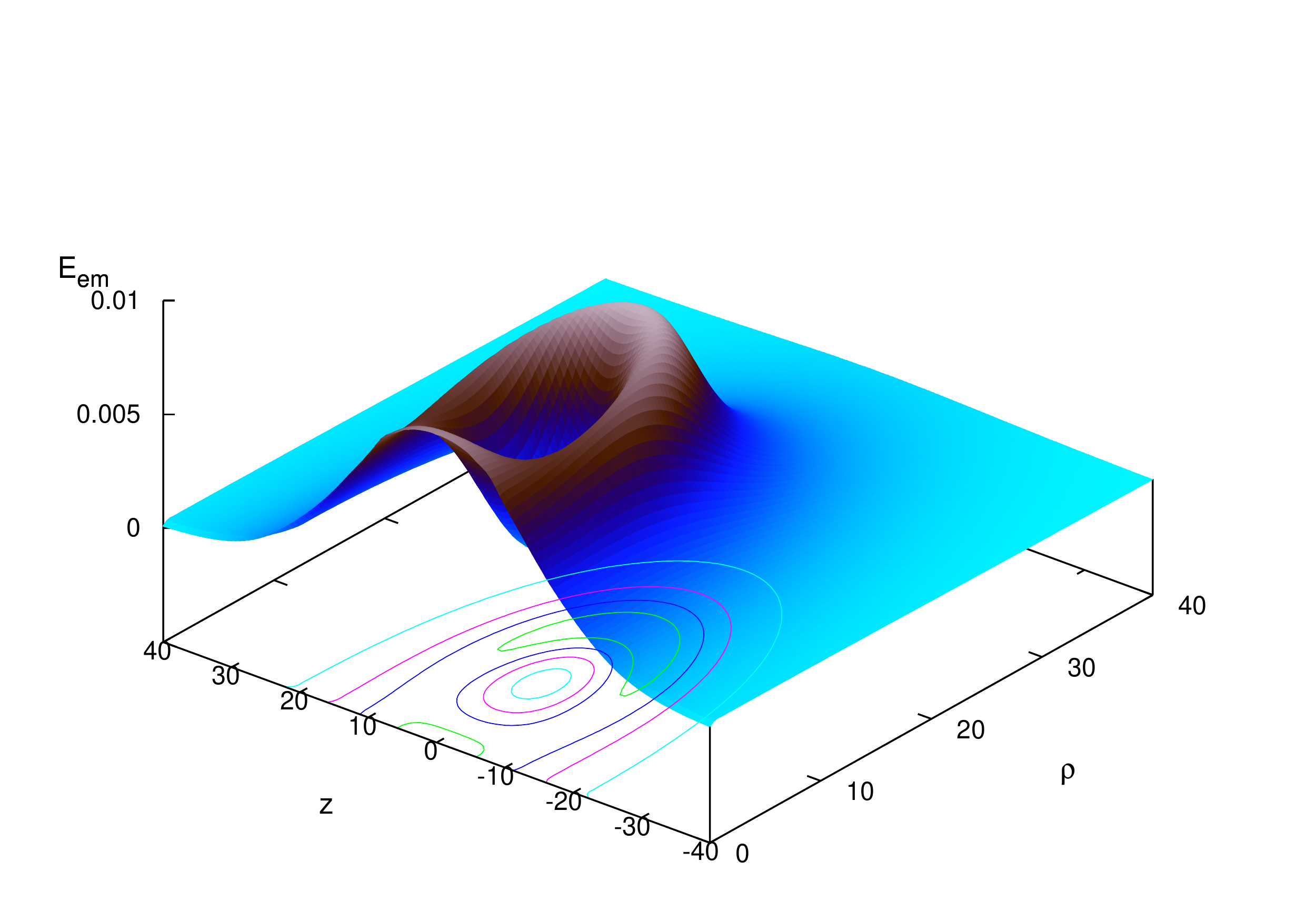}\\[18pt]
\hspace{0mm} I \hspace{67mm} II\hspace{27mm}
    \includegraphics[width=.34\textwidth, angle=-90]{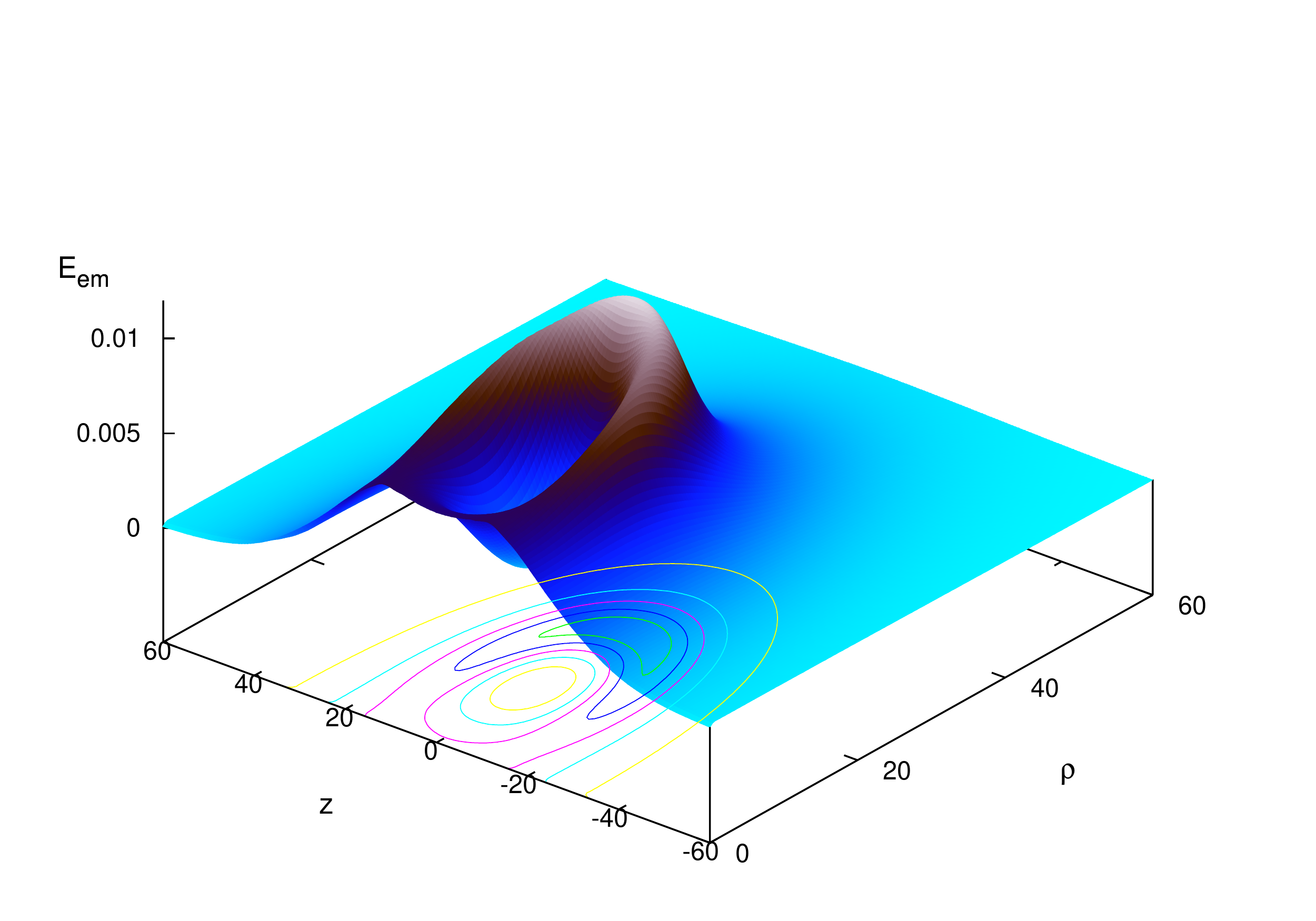}
    \includegraphics[width=.34\textwidth, angle=-90]{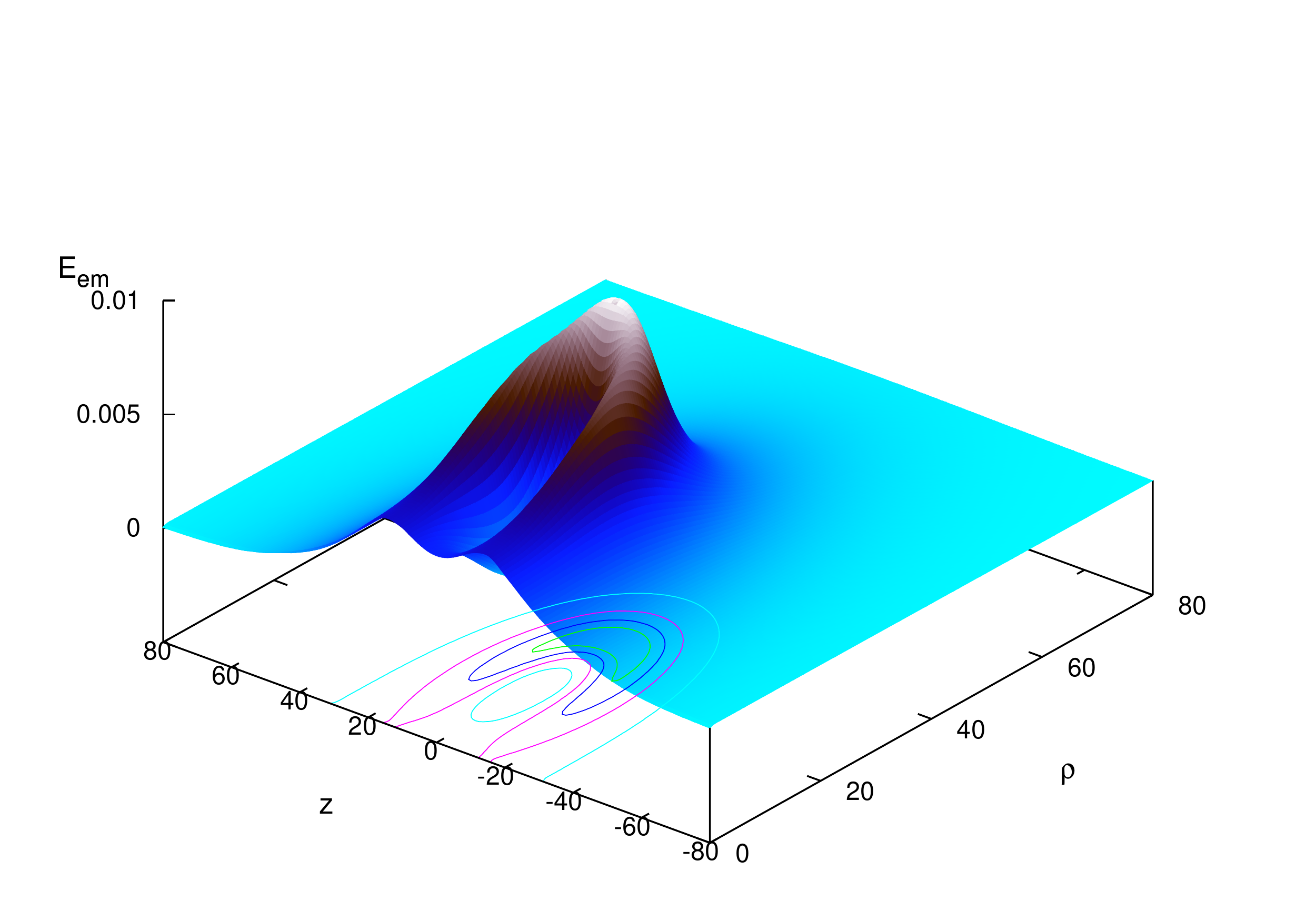}\\[18pt]
    \hspace{0mm}  III\hspace{67mm} IV\hspace{27mm}\\
    \end{center}
    \caption{\small
        The distributions of the electromagnetic energy density of the gauged parity-even Q-balls on
        four different branches at $n=1, \mu = 0.01$ and $\omega=0.60$ are shown as functions of the coordinates $\rho=r \sin\theta$
        and $z=r\cos \theta$.}
    \lbfig{Em-3d}
\end{figure}

The situation is different for the spinning gauged Q-balls with long-range component $X(r,\theta)$.
This corresponds to the case of vanishing potential
$U(\psi)$  \re{Pot}, however, the vacuum expectation value of the real
massless scalar field still $\psi$ remains nonzero  \cite{Levin:2010gp}.
We observe that in the limit $\mu \to 0$
the magnetic branch disappear and both the energy and the charge of the configuration
diverge at some critical minimal value of the angular frequency $\omega_\mathrm{min}$,
as shown in Fig.~\ref{E-omega}, left plot. The minimal critical value of the frequency
increases with the gauge coupling.

\section{Conclusions}

Our investigation confirms the existence of new type of
axially-symmetric solutions of the $U(1)$ gauged Friedberg-Lee-Sirlin model.
They exhibit examples of the configurations with both the
electric charge and toroidal magnetic field, which forms a vortex encircling
the configuration. These gauged Q-balls possess a quantized angular momentum, $J=nQ$.
We observe that the gauged Q-balls exist for relatively small values of the gauge coupling,
increase of the coupling yields stronger electromagnetic repulsion which makes the configuration unstable.
Addressing the frequency dependence of the stationary rotating Q-balls
we we found that the solutions exist only in a  frequency
range, which is restricted from  below by some critical frequency $\omega_{min}$. The value of $\omega_{min}$
depends on the strength of the gauge coupling. A novel feature of the gauged axially symmetric Q-balls is that
the corresponding branch structure is different from the ordinary Q-balls, a new magnetic branch arises at
$\omega_{min}$, it extends forward as the frequency increases. The contribution of the magnetic energy is dominating
along this branch, strong magnetic field of the vortex destroys the superconductive phase in some region
inside the Q-ball. To our best knowledge, such vorton type solutions have not been reported in the literature before.

The work here should be taken further by considering the axially symmetric gauged Q-balls in the single component
model with sextic potential \cite{Volkov:2002aj,Kleihaus:2005me,Kleihaus:2007vk,Radu:2008pp}. It
is intriguing to find in this model the solutions, which represent magnetic Q-balls, and investigate their properties.
Another interesting direction is to investigate the axially symmetric, rotating magnetic boson stars and
corresponding hairy black holes, presence of the toroidal magnetic field may lead to new interesting phenomena, in particular
in astrophysics and cosmology.

Finally, let us note that on the spacial asymptotic the system of dynamical equations \re{eqs} with the
boundary conditions \re{boundary_r} is reduced to the standard harmonic equations, both for the scalar fields and
for the components of the vector potential. Thus, by analogy with the ordinary Q-balls
\cite{Volkov:2002aj,Kleihaus:2005me,Kleihaus:2007vk}, one
can expect existence of two types of solutions, possessing different parity. In the present paper we restricted
our consideration to the $n=1$ parity-even gauged Q-balls, we hope to address the systematic study of parity-odd
solutions of the gauged Friedberg-Lee-Sirlin model in our future work.

{\bf Acknowledgements}-- We are grateful to Burhard Kleihaus, Jutta Kunz, Ilya Perapachka  and Eugen Radu
for inspiring and valuable discussions. Ya.S. would like to acknowledge useful discussions with
Emin Nugaev at the first stages of this research.
Ya.S. gratefully acknowledges the support of the Alexander
von Humboldt Foundation and from the Ministry of Education and Science of Russian Federation, project No
3.1386.2017. He would like to thank Jutta Kunz for kind hospitality at the
Department of Physics, Carl von Ossietzky University of Oldenburg
during the completion of this work.

\begin{small}

  \end{small}


\begin{thebibliography}{99}
\bibitem{Manton:2004tk}
  N.~S.~Manton and P.~Sutcliffe,
  {\it 'Topological solitons',}
    Cambridge University Press, 2004.
\bibitem{Lee:1991ax}
  T.~D.~Lee and Y.~Pang,
  Phys.\ Rept.\  {\bf 221} (1992) 251
\bibitem{Shnir2018}
Y.M.~Shnir,
{\it 'Topological and Non-Topological Solitons in Scalar Field Theories'},
Cambridge University Press, 2018.
\bibitem{Rosen}G.~Rosen, J.\ Math.\ Phys.\ {\bf 9} (1968) 996, 999
\bibitem{Friedberg:1976me}
  R.~Friedberg, T.~D.~Lee and A.~Sirlin,
  Phys.\ Rev.\ D {\bf 13} (1976) 2739
\bibitem{Coleman:1985ki}
  S.~R.~Coleman,
  Nucl.\ Phys.\ B {\bf 262} (1985) 263
   Erratum: [Nucl.\ Phys.\ B {\bf 269} (1986) 744].
\bibitem{Levin:2010gp}
  A.~Levin and V.~Rubakov,
  Mod.\ Phys.\ Lett.\ A {\bf 26} (2011) 409.
\bibitem{Loiko:2018mhb}
  V.~Loiko, I.~Perapechka and Y.~Shnir,
  Phys.\ Rev.\ D {\bf 98} (2018) no.4,  045018
\bibitem{Kaup:1968zz}
  D.~J.~Kaup,
  Phys.\ Rev.\  {\bf 172} (1968) 1331.
\bibitem{Ruffini:1969qy}
  R.~Ruffini and S.~Bonazzola,
  Phys.\ Rev.\  {\bf 187} (1969) 1767.
\bibitem{Friedberg:1986tp}
  R.~Friedberg, T.~D.~Lee and Y.~Pang,
  Phys.\ Rev.\ D {\bf 35} (1987) 3640
\bibitem{Friedberg:1986tq}
  R.~Friedberg, T.~D.~Lee and Y.~Pang,
  Phys.\ Rev.\ D {\bf 35}, 3658 (1987).
\bibitem{Kleihaus:2005me}
  B.~Kleihaus, J.~Kunz and M.~List,
  Phys.\ Rev.\ D {\bf 72} (2005) 064002
\bibitem{Kleihaus:2007vk}
  B.~Kleihaus, J.~Kunz, M.~List and I.~Schaffer,
  Phys.\ Rev.\ D {\bf 77} (2008) 064025
\bibitem{Brihaye:2008cg}
  Y.~Brihaye and B.~Hartmann,
  Phys.\ Rev.\ D {\bf 79} (2009) 064013
\bibitem{Kunz:2019sgn}
  J.~Kunz, I.~Perapechka and Y.~Shnir,
  arXiv:1904.13379 [gr-qc].
\bibitem{Jetzer:1991jr} P.~Jetzer,
Phys.\ Rept.\  {\bf 220}  (1992) 163.
\bibitem{Lee:1986ts}T.~D.~Lee,
 Phys.\ Rev.\ D  {\bf 35}  (1987) 3637.
\bibitem{Affleck}I.~Affleck and M.~Dine,
Nucl.\ Phys.\ B  {\bf 249}  (1985)  361.
\bibitem{Kusenko:1997si}A.~Kusenko and M.~E.~Shaposhnikov,
Phys.\ Lett.\ B  {\bf 418}  (1998)  46.
\bibitem{Lee:1988ag}
  K.~M.~Lee, J.~A.~Stein-Schabes, R.~Watkins and L.~M.~Widrow,
  Phys.\ Rev.\ D {\bf 39} (1989) 1665.
\bibitem{Lee:1991bn}
  C.~H.~Lee and S.~U.~Yoon,
  Mod.\ Phys.\ Lett.\ A {\bf 6} (1991) 1479.
\bibitem{Witten:1984eb}
  E.~Witten,
  Nucl.\ Phys.\ B {\bf 249} (1985) 557.
\bibitem{Copeland:1987th}
  E.~J.~Copeland, N.~Turok and M.~Hindmarsh,
  Phys.\ Rev.\ Lett.\  {\bf 58} (1987) 1910.
\bibitem{Davis:1988jp}
  R.~L.~Davis and E.~P.~S.~Shellard,
  Phys.\ Lett.\ B {\bf 207} (1988) 404.
\bibitem{Davis:1988ij}
  R.~L.~Davis and E.~P.~S.~Shellard,
  Nucl.\ Phys.\ B {\bf 323} (1989) 209.
\bibitem{Garaud:2013iba}
  J.~Garaud, E.~Radu and M.~S.~Volkov,
  Phys.\ Rev.\ Lett.\  {\bf 111} (2013) 171602
\bibitem{Kusenko:1997vi}
  A.~Kusenko, M.~E.~Shaposhnikov and P.~G.~Tinyakov,
  Pisma Zh.\ Eksp.\ Teor.\ Fiz.\  {\bf 67} (1998) 229
   [JETP Lett.\  {\bf 67} (1998) 247]
\bibitem{Anagnostopoulos:2001dh}
  K.~N.~Anagnostopoulos, M.~Axenides, E.~G.~Floratos and N.~Tetradis,
  Phys.\ Rev.\ D {\bf 64} (2001) 125006
\bibitem{Gulamov:2015fya}
  I.~E.~Gulamov, E.~Y.~Nugaev, A.~G.~Panin and M.~N.~Smolyakov,
  Phys.\ Rev.\ D {\bf 92} (2015) no.4,  045011
\bibitem{Gulamov:2013cra}
  I.~E.~Gulamov, E.~Y.~Nugaev and M.~N.~Smolyakov,
  Phys.\ Rev.\ D {\bf 89} (2014) no.8,  085006
\bibitem{Panin:2016ooo}
  A.~G.~Panin and M.~N.~Smolyakov,
  Phys.\ Rev.\ D {\bf 95} (2017) no.6,  065006
\bibitem{Loginov:2019sqf}
  A.~Y.~Loginov and V.~V.~Gauzshtein,
  Phys.\ Rev.\ D {\bf 99} (2019) no.6,  065011
\bibitem{Shiromizu:1998eh}
  T.~Shiromizu,
  Phys.\ Rev.\ D {\bf 58} (1998) 107301
\bibitem{Volkov:2002aj}M.S.~Volkov and E.~Wohnert,
Phys.\ Rev.\  D  {\bf 66} (2002)  085003.
\bibitem{Radu:2008pp}E.~Radu and M.S.~Volkov,
Phys.\ Rept.\  {\bf 468}  (2008)  101.
\bibitem{schoen}
W.~Sch\"onauer  and R.~Wei\ss,
"Efficient vectorizable PDE solvers"
J. Comput. Appl. Math. 1989. V. 27. P. 279\\
M.~Schauder, R.~Wei\ss and  W.~Sch\"onauer,
"The CADSOL Program Package",
Universit\"at Karlsruhe, 1992. Interner Bericht Nr. 46/92.
\bibitem{Tamaki:2010zz}
  T.~Tamaki and N.~Sakai,
  Phys.\ Rev.\ D {\bf 81} (2010) 124041
\bibitem{Kleihaus:2011sx}
  B.~Kleihaus, J.~Kunz and S.~Schneider,
  Phys.\ Rev.\ D {\bf 85} (2012) 024045
\bibitem{Nugaev:2019vru}
  E.~Nugaev and A.~Shkerin,
  arXiv:1905.05146 [hep-th].


\end{thebibliography}
\end{document}